\documentclass[letterpaper,12pt]{article}

\usepackage[margin=1in]{geometry}
\usepackage{amsmath,amssymb,amsthm,mathrsfs,mathtools}
\usepackage{bm}
\usepackage{setspace}
\usepackage{fancyhdr}
\usepackage{url}
\usepackage{color,graphicx}
\usepackage{qtree}
\usepackage{wrapfig}
\usepackage{caption}
\usepackage{subcaption}
\usepackage[linesnumbered,ruled,vlined]{algorithm2e}
\usepackage{tabularx}
\usepackage{xr}
\makeatletter
\newcommand*{\addFileDependency}[1]{
  \typeout{(#1)}
  \@addtofilelist{#1}
  \IfFileExists{#1}{}{\typeout{No file #1.}}
}
\makeatother

\newcommand*{\myexternaldocument}[1]{%
    \externaldocument{#1}%
    \addFileDependency{#1.tex}%
    \addFileDependency{#1.aux}%
}
\myexternaldocument{supp}

\usepackage{lineno}
\usepackage[numbers]{natbib}
\usepackage{tikz}
\usetikzlibrary{arrows,chains,positioning,scopes,calc,decorations.pathreplacing}
\tikzset{
  treenode/.style = {align=center, inner sep=0pt, text centered,
    font=\sffamily},
  arn_n/.style = {treenode, font=\sffamily\bfseries, draw=black,
    text width=1.5em},
  arn_r/.style = {treenode, circle, black, draw=black, 
    text width=1.5em},
  arn_y/.style = {treenode, circle,  black, draw=black, text width=1.5em, fill = block-gray}
}
\usepackage[most]{tcolorbox}
\definecolor{block-gray}{gray}{0.85}
\newtcolorbox{blockquote}{colback=block-gray,grow to right by=-1mm,grow to left by=-1mm,boxrule=0pt,boxsep=0pt,breakable}

\def\bone{{\bf 1}}
\def\bee{\begin{equation}}
\def\eed{\end{equation}}
\def\codacore{{CoDaCoRe}}

\def\clr{{\rm clr}}

\def\bx{{\boldsymbol x}}
\def\by{{\boldsymbol y}}
\def\bz{{\boldsymbol z}}

\def\bX{{\bf X}}

\def\thick#1{\hbox{\rlap{$#1$}\kern0.25pt\rlap{$#1$}\kern0.25pt$#1$}}
\def\balpha{\boldsymbol{\alpha}}
\def\bbeta{\boldsymbol{\beta}}
\def\bgamma{\boldsymbol{\gamma}}

\def\smbalpha{\boldsymbol{{\scriptstyle{\alpha}}}}

\def\betahat{{\widehat\beta}}

\def\sigmahat{{\widehat\sigma}}

\def\psihat{{\widehat\psi}}

\def\smbalpha{\widehat{\smbalpha}}

\def\hbar{\bar{ h}}

\def\zbar{\bar{ z}}

\def\bybar{\bar{ \by}}

\def\Ssc{{\cal S}}

\def\real{{\mathbb R}}

\def\transpose{{\sf \scriptscriptstyle{T}}}

\def\trans{^{\transpose}}

\def\mybox#1{\vskip1mm \begin{center}
        \hspace{.0\textwidth}\vbox{\hrule\hbox{\vrule\kern6pt
\parbox{.9\textwidth}{\kern6pt#1\vskip6pt}\kern6pt\vrule}\hrule}
        \end{center} \vskip-5mm}
\def\lboxit#1{\vbox{\hrule\hbox{\vrule\kern6pt
      \vbox{\kern6pt#1\vskip6pt}\kern6pt\vrule}\hrule}}

\def\thickboxit#1{\vbox{{\hrule height 1mm}\hbox{{\vrule width 1mm}\kern6pt
          \vbox{\kern6pt#1\kern6pt}\kern6pt{\vrule width 1mm}}
               {\hrule height 1mm}}}

\def\fat#1{\hbox{\rlap{$#1$}\kern0.25pt\rlap{$#1$}\kern0.25pt$#1$}}

\theoremstyle{remark}

\def\bee{\begin{equation}}
\def\eed{\end{equation}}
\def\codacore{{CoDaCoRe}}

\providecommand{\keywords}[1]
{
  \small	
  \textbf{\textit{Keywords:}} #1
}

\title{Variable selection in balance regression with applications to microbiome compositional data}
\author{Jing Ma\footnote{Correspondence: {jingma@fredhutch.org}}\\
Division of Public Health Sciences, Fred Hutchinson Cancer Center
 \and
 Paizhe Xie\\
 Department of Statistics, University of Washington
 \and
 Kristyn Pantoja\footnote{This work was partially completed while KP was a graduate student at Texas A\&M University.}\\
 Novartis
 \and
 David E. Jones\\
Department of Statistics, Texas A\&M University
}

\usepackage{placeins}

\begin{document}

\maketitle



\begin{abstract}
Compositional data, where only relative abundances are available, are common in microbiome and other high-throughput sequencing studies. Log ratios between groups of variables serve as key biomarkers in these settings. However, selecting predictive log ratios is a combinatorial challenge, and existing greedy search-based methods are computationally expensive, limiting their applicability to high-dimensional data. We introduce the supervised log ratio (SLR) method, a novel and efficient approach for selecting predictive log ratios in high-dimensional settings. SLR first screens active variables using univariate regression on log ratio transformed data and then applies principal balance analysis to define balance biomarkers. Our approach leverages both the relationship between the response and predictors and the correlations among the predictors to improve accuracy in variable selection and prediction. Through simulations and two case studies—one on inflammatory bowel disease (IBD) and another on colorectal cancer (CRC)—we demonstrate that SLR outperforms existing methods, particularly in high-dimensional settings. SLR is implemented in an R package, publicly available at \url{https://github.com/drjingma/slr}. 

\end{abstract}

\doublespacing

\section{Introduction}

In microbiome profiling studies, measuring the absolute abundance of each taxon is often costly \citep{barlow2020quantitative}. However, feature relative abundances can be easily obtained through shotgun metagenomics or 16S rRNA sequencing, resulting in compositional data \citep{gloor2017microbiome}. A key goal of these studies is to identify interpretable biomarkers that can predict health outcomes \citep{sepich2021microbiome}. 

\emph{A compositional balance}---defined as the log ratio between the geometric means of two groups of taxa \citep{egozcue2005groups}---serves as such a biomarker. Ratios overcome the challenge of not knowing the absolute abundances \citep{gloor2017microbiome,morton2019establishing} and provide robustness against sequencing-induced multiplicative bias \citep{mclaren2019consistent}. Log ratios enforce symmetry around zero and preserve key principles of compositional data analysis, such as scale-invariance. As a result, balances offer a meaningful scalar summary of microbiome compositions and can be used for diagnosis, prognosis, or predicting therapeutic responses \citep{morton2017balance,silverman2017phylogenetic}. Moreover, each of the two taxon groups that form a balance represents a symbiotic consortium, either beneficial or pathogenic, which could serve as a foundation for microbiome-based therapeutics \citep{atarashi2013treg,sorbara2022microbiome,van2021rationally,louie2023ve303}. Indeed, research suggests that microbial consortia can be more effective than single strains \citep{atarashi2013treg}, as they exhibit greater stability and adaptability due to interspecies interactions \citep{cao2022construction} and division of labor \citep{duncker2021engineered}.  

Balances can be defined with respect to a pre-specified sequential binary partition (SBP) \citep{morton2017balance,silverman2017phylogenetic}. However, when such SBPs are unavailable or not applicable, selecting balance biomarkers is a combinatorial problem. Selbal \citep{rivera2018balances} starts with pairwise log-ratios and adds one best variable at a time to search for the optimal balance. Unfortunately, their method does not guarantee the global optimum because it does not search all possible balances. Furthermore, their method is computationally prohibitive in high-dimensional settings. More recently, \citep{gordon2022learning} proposed {\codacore} which approximates the underlying combinatorial problem with a continuous relaxation. The resulting procedure is computationally more efficient than selbal. Besides selbal and {\codacore}, codaLasso is another popular method for predictive modeling of compositional data. It is based on the linear log-contrast model \citep{aitchison1984log} and uses Lasso penalization to perform variable selection  \citep{lin2014variable,shi2016regression,lu2019generalized}. Although codaLasso is not strictly a balance regression method, its fitted regression model can be interpreted as a weighted balance between two groups of taxa \citep{susin2020variable}. 

The performance of the above-mentioned methods in classification of a health outcome has been well studied \citep{gordon2022learning,susin2020variable,quinn2020interpretable}. However, their comparative performance in variable selection is not fully understood. This is disappointing especially given that they were developed with the goal of selecting interpretable biomarkers. Understanding the strengths and weaknesses of existing balance regression methods in variable selection, especially in high-dimensional settings, is essential to ensure the reliability of biological conclusions made on the basis of applying these methods. 

We bridge this gap with two key contributions. First, we conduct data-driven simulations using two studies---one on the inflammatory bowel diseases (IBD) and the other on colorectal cancer (CRC)---to benchmark the performance of selbal, CoDaCoRe, and codaLasso. We demonstrate that existing balance regression methods perform poorly when the features to be selected are highly correlated. High correlation among features can arise when a group of taxa collectively contribute to the same biological process or exploit the same class of resources in a similar way. These taxa are sometimes referred to as an ecological ``guide" \citep{wu2021guild}. 
Second, we introduce a new approach called the {\it supervised log ratio} (SLR) method for high-dimensional balance regression. SLR consists of two main steps. First, it screens for active variables by performing univariate regression of the response on each taxon after a centered log ratio transformation. Next, SLR uses the leading principal balance \citep{martin2018advances} on the reduced data to define a balance biomarker. SLR can be viewed as a semi-supervised approach: it leverages the response variable to filter out inactive features while using principal balance analysis to capture latent structures in the data. This distinguishes it from PLS-PB \citep{nesrstova2023principal}, which derives interpretable balances using partial least squares. However, PLS-PB does not lead to sparse loadings and is not suitable for variable selection. Our simulations confirm that SLR improves both the sensitivity and stability of variable selection compared to existing methods. Additionally, SLR outperforms existing methods in terms of classification accuracy and variable selection on the IBD and CRC data sets. Finally, SLR is computationally efficient as it only requires univariate regression and a simple modification of principal component analysis.


\section{Methods}\label{sec:method}

\subsection{Data sets}

We downloaded two data sets from Borenstein Lab's GitHub repository \url{https://github.com/borenstein-lab/microbiome-metabolome-curated-data} \citep{muller2022gut}. 

The first data set comes from \citep{franzosa2019gut} which consists of stool samples from a discovery and a validation cohort of ulcerative colitis (UC) and non-IBD controls. There are 34 controls and 53 UC subjects in the discovery cohort, and 22 controls and 23 UC subjects in the validation cohort. The studies in \citep{franzosa2019gut} also included Crohn's disease (CD) patients. {We focused only on UC because the metagenomic profiles of UC patients are less distinct from the profiles of the controls compared to CD patients \citep{franzosa2019gut}, which makes it more suitable for differentiating all the methods considered.} 

The second data set comes from \citep{yachida2019metagenomic} which has stool samples from 127 healthy subjects and 123 CRC patients. These patients were in different cancer stages (I, II, III and IV), but we pooled them together as a single group. We did not include subjects with polyps only or who had a history of colorectal surgery but were otherwise normal. We also excluded subjects in stage 0 CRC. 

Both data sets were obtained via shotgun metagenomic sequencing. The minimum sequencing depth per sample is 126,087 in the IBD data set and 2,892,954 in the CRC data set. For each data set, we removed taxa whose mean relative abundance is below $5\cdot 10^{-5}$. Thresholding by mean abundance removes most of rare taxa. For the CRC data set, we also filtered taxa whose prevalence is below 30\%. The resulting data matrix has 394 taxa and about 3\% zeros. For the IBD data set, we filtered taxa whose prevalence is below 50\%. 
The resulting data matrices contain 447 taxa, with 4\% zeros. 
A smaller threshold for the mean relative abundance and/or prevalence could be used, which would lead to more zeros in the data and even higher dimensionality. However, since methods for balance regression require strictly positive relative abundances, an overrepresentation of zeros would violate their model assumptions and make it difficult to fairly evaluate their performance. 

Before proceeding with balance regression, all zeros in the data were imputed using the geometric Bayesian multiplicative (GBM) replacement \citep{martin2015bayesian} as implemented in the `cmultRepl()' function of the R package {\tt zCompositions} \citep{palarea2015zcompositions}. Other zero replacement strategies can also be used. 

\subsection{Simulation}

We simulated separate training and test cohorts for each data set. Since the CRC data set does not have a test cohort, we partitioned this data set into a training cohort of 175 subjects and a test cohort of 75 subjects randomly.

To generate bacterial consortia, we used the leading eigenvectors of the empirical correlation matrix of the full data (training set only) after the centered log ratio (CLR) transformation. Given an eigenvector $V$, the active features were taken as the ones with the top 1\% largest values ($I_{+}$) and the ones with the bottom 1\% smallest values ($I_{-}$) in $V$. Given the active features, we used the following steps to generate the data. First, we randomly sampled $d$ inactive features ($I_0$) from the full feature list excluding those already included as active features ($I_{+} \cup I_-$). Second, the training ($X$) and test ($X^{\rm test}$) data were obtained by extracting columns of the respective full data matrices corresponding to the list of features used ($I_{+} \cup I_- \cup I_0$). Third, the balance $z$ ($z^{\rm test}$) is computed based on the design matrix $X$ ($X^{\rm test}$) and the partition defined by $I_{+}, I_-, $ and $I_0$. Lastly, we sampled binary responses from the Bernoulli distribution $y_i = \mbox{Bernoulli}(\pi_i)$ and $y^{\rm test}_i = \mbox{Bernoulli}(\pi^{\rm test}_i)$ with $ \pi_i = 1/(e^{-b_0 - b_1 z_i}+1)$ and $ \pi^{\rm test}_i = 1/(e^{-b_0 - b_1 z_i^{\rm test}}+1)$, where $b_1=1$. Denote by $\zbar$ the mean of $z$. We used $b_0 = \log(53/34) - b_1 \zbar$ in the IBD data set, 
and $b_0 = \log(123/127) - b_1 \zbar$ in the CRC data set. We chose these intercepts so that the distribution of the responses is approximately the same as that in the original data. It is worth noting that once the partition is determined, the design matrices $X$ and $X^{\rm test}$ become fixed. The only source of noise comes from the response model. 

In the simulated IBD data, the number of active features is 10, while the number of inactive features $d\in \{50,400\}$. In the simulated CRC data, the number of active features is 8 and the number of inactive features $d\in \{50,375\}$. The use of different $d$'s is to evaluate the impact of dimensionality on different methods. All comparisons were evaluated with 50 replications.  We used area under the curve (AUC) for classification error on the test data set and sensitivity and specificity for variable selection accuracy. For a partition $\beta$ and its estimate $\betahat$, sensitivity and specificity are defined, respectively, as
\[
{\rm sensitivity} = \frac{|\{j: \betahat_j \ne 0, \beta_j \ne  0\}|}{|\{j: \beta_j \ne 0\}|}, \quad
{\rm specificity} =  \frac{|\{j: \betahat_j = 0, \beta_j =  0\}|}{|\{j: \beta_j = 0\}|}.
\] 
We also recorded the percentage of variables selected and the computational time on a Linux machine that had an Intel Core i9 processor with 18 cores (36 threads) and 128 GB memory (inclusive of time used for model selection).

\subsection{Balance regression}

Let $\Ssc^p=\{(x_1, \ldots, x_p)\trans: x_j \ge 0\ (j=1,\ldots,p),~ x_1+\ldots+x_p =1\}$ denote the $p$-dimensional simplex. Let $\bX=(X_1,\ldots,X_p)\trans\in \Ssc^p$ be a $p$-dimensional compositional vector. The balance between variables from two non-overlapping subsets $I_{+}\subset\{1,\ldots,p\}$ and $I_{-}\subset\{1,\ldots,p\}$ is defined as
\begin{equation}\label{e:bade}
B(\bX; I_{+},I_{-}) = \sqrt{\frac{|I_{+}||I_{-}|}{|I_{+}|+|I_{-}|}} \log \frac{g(\bX_{I_{+}})}{g(\bX_{I_{-}})} \propto \frac{1}{|I_{+}|} \sum_{j \in I_{+}} \log X_j - \frac{1}{|I_{-}|} \sum_{j \in I_{-}} \log X_j,
\end{equation}
where $|I|$ denotes the size of the subset $I$, $\bX_{I}$ the sub-vector of $\bX$ whose elements are the variables indexed by $I$, and the sign $\propto$ means proportional. From a dimensionality reduction perspective, a balance provides a meaningful scalar summary of a compositional vector, leading to easier interpretation and translational use as a clinical biomarker. 

Suppose we have $n$ independent and identically distributed observations $(\bx_i, y_i)$, where $\bx_i = (x_{i,1}, \ldots, x_{i,p})\trans \in \Ssc^p$ is a vector of relative abundances and $y_i \in \mathbb{R}$ is a response variable, for $i=1,\ldots,n$. 
For a response $y$ and its linear predictor $f\in \real$, denote by $L(y,f) = (y-f)^2$ the sum-of-squares loss if $y$ is continuous and $L(y,f) = y f - A(f)$ with $A(f) = \log (1 + e^f)$ if $y$ is dichotomous. 
Balance regression \citep{rivera2018balances} seeks to find the best subsets $I_{+}$ and $I_{-}$ that solve the following optimization problem 
\begin{equation}\label{e:bare}
\min_{I_+, I_-, \theta_0, \theta_1} \left\{ \sum_{i=1}^n L(y_i, \theta_0 + \theta_1 B(\bx_i; I_{+},I_{-}))\right\}.
\end{equation}
Without loss of generality, we assume $\theta_1\ge 0$ so that $I_+$ includes variables that are positively associated with the response. It is straightforward to expand Equation \eqref{e:bare} into the linear log contrast model with coefficients $\bbeta$ defined as $\beta_j = \theta_1 \sqrt{{|I_{-}|}/\{|I_{+}|({|I_{+}|+|I_{-}|})\}}$ for $j\in I_+$, $\beta_j = -{\theta_1} \sqrt{{|I_{+}|}/\{|I_{-}|({|I_{+}|+|I_{-}|})\}}$ for $j\in I_-$, and zero elsewhere. {While the coefficients in a balance regression model are restricted to have only two distinct values, the flip side is this leads to a simpler model which, if estimated more accurately, can lead to gain in prediction performance and variable selection.}

\subsection{Supervised log ratios}
\emph{Supervised Log Ratio} or SLR, is a method for balance selection built on two simple ideas: feature screening and principal balance analysis. Feature screening has long been used as an effective approach for dimensionality reduction \citep{fan2008sure,zhu2011model,fan2009ultrahigh}. For example, it has been used in combination with principal component analysis (PCA) or clustering to analyze gene expression data \citep{hastie2000gene,bair2004semi,bair2006prediction}. However, interpretation of PCA loadings can be challenging because they typically involve all variables. By contrast, principal balance analysis (PBA) \citep{pawlowsky2011principal,martin2018advances} provides loadings that can be interpreted as balances. Still, by itself PBA is an unsupervised method. Recently, \citep{nesrstova2023principal} introduced a supervised method PLS-PB which combines PBA with partial least squares for interpretable dimensionality reduction. Unfortunately, PLS-PB does not lead to parsimonious selection of biomarkers. By contrast, SLR is a supervised method that also provides feature selection. 

Suppose $y_i \in \mathbb{R}$ is a continuous response variable, for $i=1,\ldots,n$. We discuss other types of  response variables later. The algorithm for SLR can be summarized as follows:
\begin{enumerate}
\item Compute the univariate regression coefficients for each taxon. 
\item Form a reduced data matrix consisting of only those taxa whose univariate coefficient exceeds a threshold $\tau\ge 0$ in absolute value. The threshold $\tau$ is chosen via CV. 
\item Compute the leading principal balance (PB) of the reduced data matrix. 
\item Fit an ordinary least squares by regressing $y_i$ onto the leading PB. 
\end{enumerate}

We now give details of the method. Given input data $\bx_i$, let $z_{i,j} = \log(x_{i,j}) - \log g(\bx_i)$ denote the {\clr}-transformed proportions, where $g(\cdot)$ is the geometric mean function. Let $\bz_{\cdot,j} = (z_{1,j},\ldots,z_{n,j})\trans$ denote the vector of observations for the $j$-th taxon and $\bybar$ the sample mean of $\by=(y_1,\ldots,y_n)\trans$. The univariate effect of the $j$-th taxon on the response can be estimated by:
\begin{equation}\label{e:univeffect}
\psihat_j = \frac{(\by-\bybar)\trans(\bz_{\cdot,j}-\bar{\bz_{\cdot,j}})}{\|\bz_{\cdot,j}-\bar{\bz_{\cdot,j}}\|^2}.
\end{equation}
Note that the scale estimate $\sigmahat$ common to all variables cancels out. Let $C_{\tau}$ be the collection of indices such that $|\psihat_j|\ge \tau$, i.e. the variables with association exceeding a threshold $\tau\ge 0$. 

We denote by $\bX_{\tau}$ the reduced data matrix consisting of columns $\bX$ restricted to indices in $C_{\tau}$. PBA of $\bX_{\tau}$ can be approached using a few different algorithms. The optimal algorithm requires an exhaustive search along all the possible partitions, which is computationally demanding \citep{martin2018advances}. We opt to a suboptimal but faster constrained PC algorithm instead \citep{chipman2005interpretable}. {Let $\bgamma_1$ be the leading PC of the {\clr}-transformed version of $\bX_{\tau}$ and $\balpha_1$ be the corresponding principal balance. The components $\alpha_{1,j}$ in $\balpha_1$ take values $\{-c_1, 0, c_2\}$ such that $\balpha_1\trans\balpha_1 = 1$ and $\balpha_1\trans \bone = 0$, where $\bone$ is a vector of ones. The constrained PC algorithm determines $\balpha_1$ by minimizing the angle between $\balpha_1$ and $\bgamma_1$, given by $\arccos(\balpha_1\trans\bgamma_1)$. The procedure begins by identifying the largest positive and negative elements in $\bgamma_1$ and assigning the corresponding elements in $\balpha_1$ the values $\pm\sqrt{2}/2$. All other elements of $\balpha_1$ are set to zero. This procedure is then repeated from 3 to $q$ elements (selected by absolute value), where $q$ is the number of taxa in $C_{\tau}$. Among these $q-1$ possible directions, the closest to $\bgamma_1$ is chosen as $\balpha_1$.} This constrained PC algorithm is much faster than the optimal algorithm with little loss in variance explained \citep{martin2018advances}. 

We treat the threshold $\tau$ as a tuning parameter. A larger $\tau$ favors a sparser model but may compromise prediction accuracy. In practice, we use CV to choose an optimal $\tau$ that balances between model sparsity and prediction performance. The candidate thresholds for $\tau$ are chosen as a length $p$ grid informed by the univariate association between each taxon and the response. 

The SLR framework can be easily extended to accommodate other types of responses, where a generalized linear model can be used to perform feature screening. Consider, for example, a binary response variable $y_i\in\{0,1\}$. In this case, the univariate effect $\psihat_j$ does not have an explicit form as in \eqref{e:univeffect}. Nonetheless, $\psihat_j$ can be estimated by fitting a simple logistic regression, e.g. using the {\tt glm} function in R.

\subsection{Model selection}

All the methods considered depend on tuning parameters. In selbal, the tuning parameter is the number of taxa included in the model. In {\codacore}, the tuning parameter is a threshold $t\in (0,1)$ used to convert the soft assignment weights associated with each taxon into hard assignments. {\codacore} chooses a grid of 20 candidate thresholds, informed by the estimated soft assignments. The tuning parameter in codaLasso is an $\ell_1$ regularization parameter that controls the sparsity of taxa. In SLR, the tuning parameter is a threshold $\tau\in(0,1)$ that controls the number of taxa included in PBA.  

For all the methods we use CV to select the optimal tuning parameters. In $K$-fold CV, the data are randomly partitioned into $K$ equal sized subsamples, also known as folds. The model is trained $K$ times, each time with one subsample used as the validation data for testing the model's prediction error and the remaining $(K-1)$ subsamples used as the training data. The average prediction error, Mean Squared Error (MSE) for continuous responses or Area Under the Curve (AUC) for dichotomous responses, across all subsamples is computed. The best model is selected as the one that minimizes the prediction error (the MIN rule). Alternatively, the ``one standard error'' (1SE) rule can be used to select the most parsimonious model whose error is no more than one standard error above the error of the best model \citep{hastie2009elements,breiman2017classification}. In practice, the 1SE rule is often recommended for simpler interpretation. Throughout the paper, all results were obtained using the 1SE rule unless otherwise mentioned.  

\subsection{Methods not included in the analysis}

There are several other approaches for predictive modeling of compositional data besides the methods presented here. For example, DBA \citep{quinn2020interpretable}, short for discriminant balance analysis, finds a sparse set of balances formed by no more than 3 taxa. Coda4microbiome \citep{calle2023coda4microbiome} finds pairwise log ratios that are predictive of a response. Coda4microbiome was introduced as a computationally efficient alternative to selbal. From a methodology perspective, coda4microbiome is similar to log ratio lasso \citep{bates2019log,fei2024scalable} which performs penalized regression on all pairwise log ratios. However, both DBA and coda4microbiome return multiple balances as supposed to a single balance. PLS-PB \citep{nesrstova2023principal} was recently developed for supervised learning by combining PBA with PLS. However, PLS-PB does not provide sparse selection of taxa. BalReg is a Bayesian balance regression method that can provide simultaneous estimation and inference \citep{huang2020bayesian}. For binary responses, BalReg employs probit regression, which relies on a latent variable. Posterior sampling for this variable requires inverting a matrix with dimensions equal to the sample size. As a result, BalReg becomes computationally demanding, even for datasets with just a few hundred observations. 

\section{Results}\label{sec:sims}

\subsection{Simulation studies}

\begin{figure}[h]
    \centering
    \includegraphics[width=0.95\linewidth]{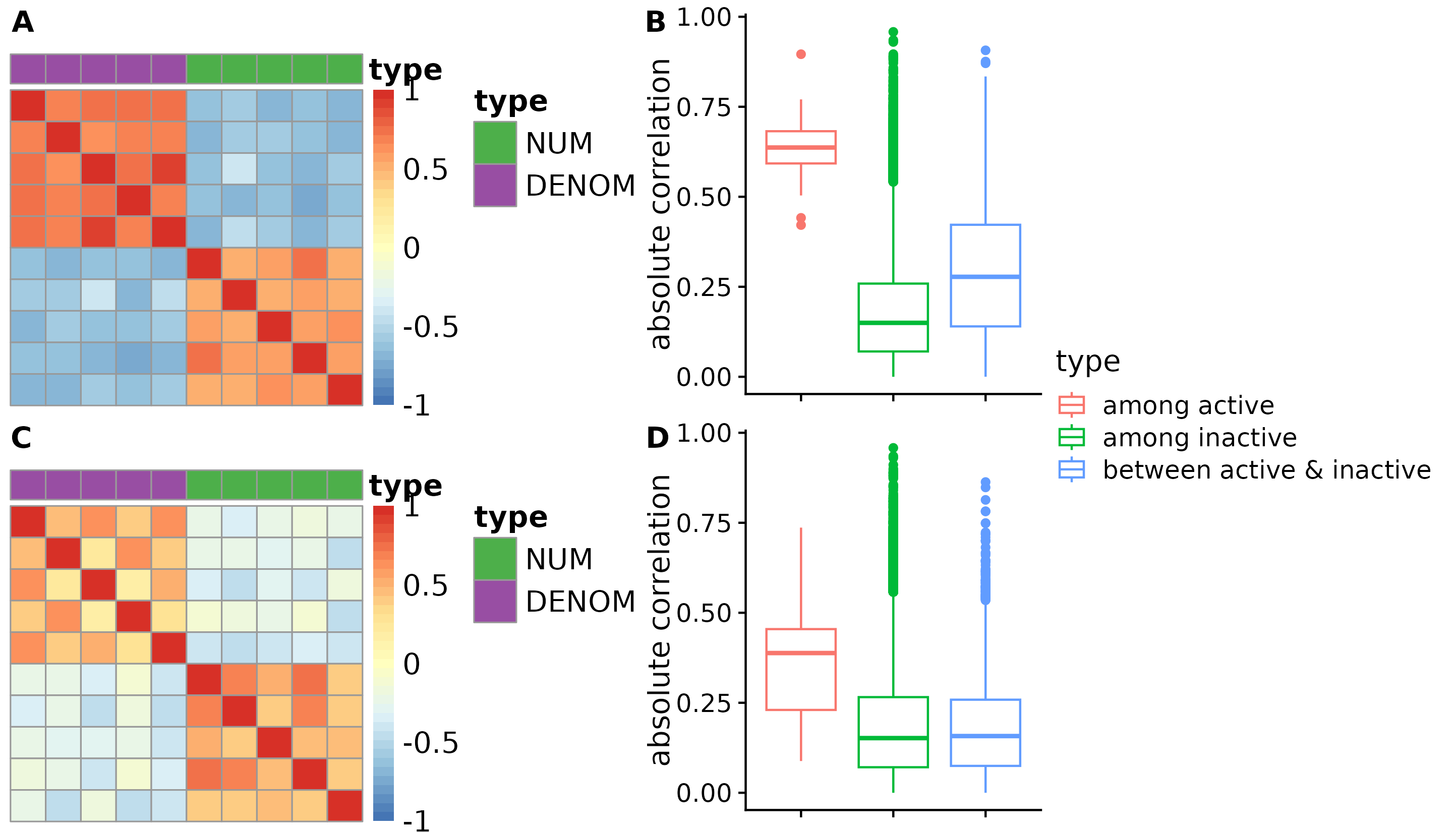}
    \caption{Correlation in the simulations based on the IBD data set ($d=400$). (A) Heat map of the correlation among active features in Scenario I. (B) Boxplots of the absolute correlation coefficients among different sets of features in Scenario I. (C) Heat map of the correlation among active features in Scenario II. (D) Boxplots of the absolute correlation coefficients among different sets of features in Scenario II.}
    \label{fig:correlation}
\end{figure}

Our simulations are designed to reflect the correlation structure expected in bacterial consortia. We considered two different scenarios. Scenario I used the top eigenvector to generate active features, while Scenario II used the third eigenvector. In both scenarios, active features form correlated groups (Figures~\ref{fig:correlation}, \ref{fig:correlation:UC50}, \ref{fig:correlation:CRC50} \& \ref{fig:correlation:CRC375}). Specifically, features within $I_+$ (the numerator group) or $I_-$ (the denominator group) are positively correlated with each other but features in the numerator group are negatively correlated with those in the denominator group (Figure~\ref{fig:correlation}A \& \ref{fig:correlation}C). This mimics the definition of bacterial consortia. A key difference between the two scenarios is that the active features in Scenario I exhibit stronger correlations than those in Scenario II (Figure~\ref{fig:correlation}B \& \ref{fig:correlation}D). This helps assess the impact of correlation structure on the performance of different methods. As shown in Figure~\ref{fig:correlation}B \& \ref{fig:correlation}D, correlation between active and inactive features can also be high. An effective method should ideally select only the active features that are directly associated with the response, while avoiding those that are merely correlated with the active features but have no direct association with the response.

\begin{figure}[h]
    \centering
    \includegraphics[width=\linewidth]{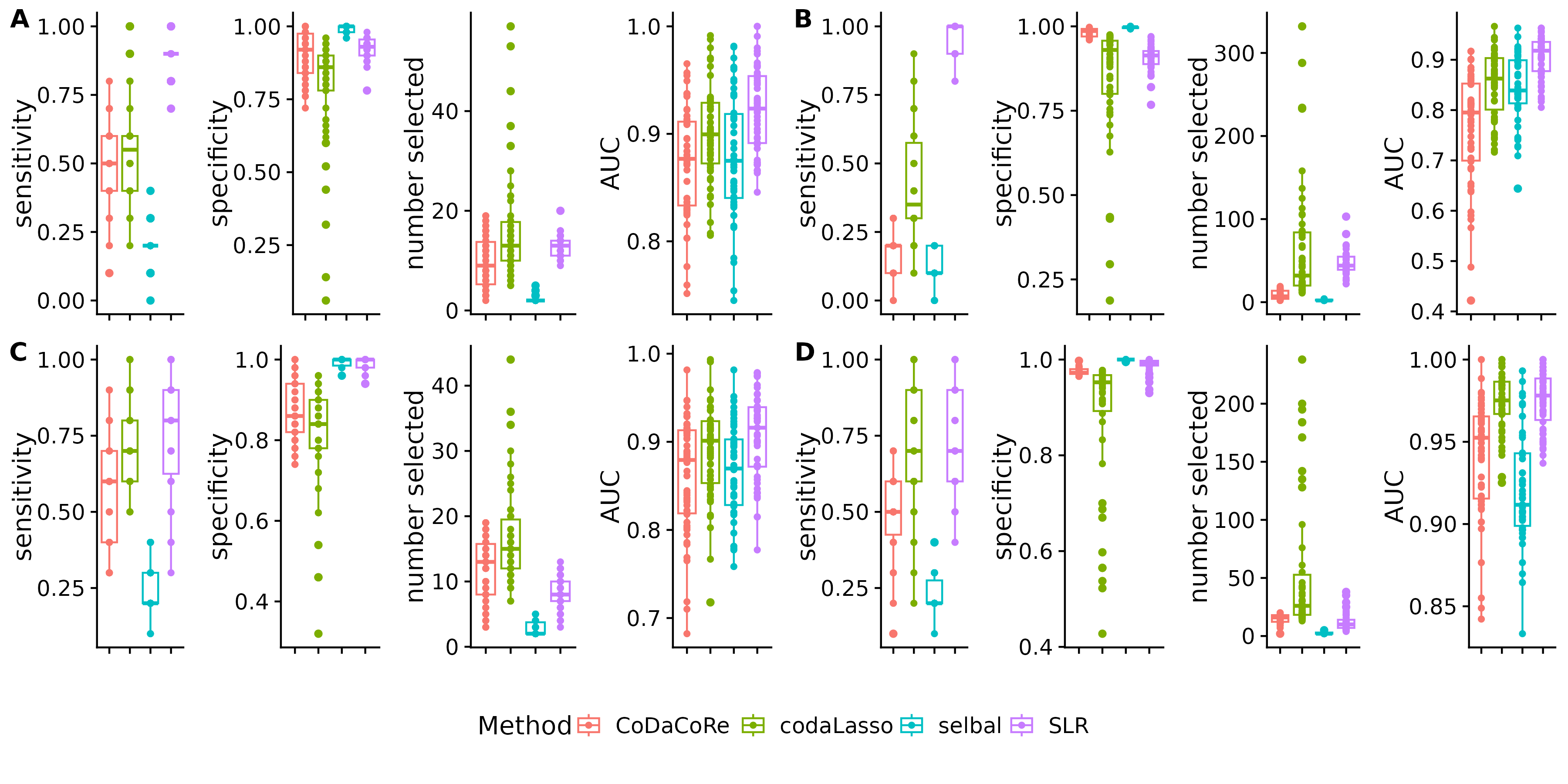} 
    \caption{Variable selection and prediction results in the simulations based on the IBD data set. Metrics used for evaluation include sensitivity, specificity, AUC, and number of variables selected. (A) Metrics in Scenario I when $d=50$. (B) Metrics in Scenario I when $d=400$. (C) Metrics in Scenario II when $d=50$. (D) Metrics in Scenario II when $d=400$.}
    \label{fig:UCsim}
\end{figure}

{We first evaluated the impact of correlation and dimensionality on variable selection and classification accuracy in the simulations based on the IBD data set. Figure~\ref{fig:UCsim} summarizes the results under different correlation structures and dimensionalities. Across all settings, selbal consistently exhibits the lowest sensitivity and highest specificity, as it tends to select very few variables. In contrast, codaLasso selects the most variables, resulting in higher sensitivity than selbal but the lowest specificity among all four methods. {\codacore} falls between selbal and codaLasso in terms of performance. Sensitivity decreases for selbal, codaLasso, and {\codacore} when the dimensionality increases, with a more pronounced drop in Scenario I, where bacterial consortia were generated using the top eigenvector (Figure~\ref{fig:UCsim}A \& \ref{fig:UCsim}B). SLR achieves the highest sensitivity in all cases except for the low-correlation, high-dimensional setting (Figure~\ref{fig:UCsim}D). While its specificity is slightly lower in Scenario I (Figure~\ref{fig:UCsim}A \& \ref{fig:UCsim}B) due to over-selection of variables, it outperforms codaLasso and {\codacore} in specificity in Scenario II (Figure~\ref{fig:UCsim}C \& \ref{fig:UCsim}D).
Notably, SLR is more robust to increasing dimensionality than the other methods but is more affected by changes in correlation. In Scenario II, its sensitivity decreases, while codaLasso and {\codacore} show improved sensitivity.
In terms of classification accuracy, SLR achieves a higher AUC than {\codacore} and selbal and sometimes outperforms codaLasso. While codaLasso is the next-best method in AUC, its advantage primarily stems from selecting a significantly larger number of variables than the other methods.}

\begin{figure}[h]
    \centering
    \includegraphics[width=\linewidth]{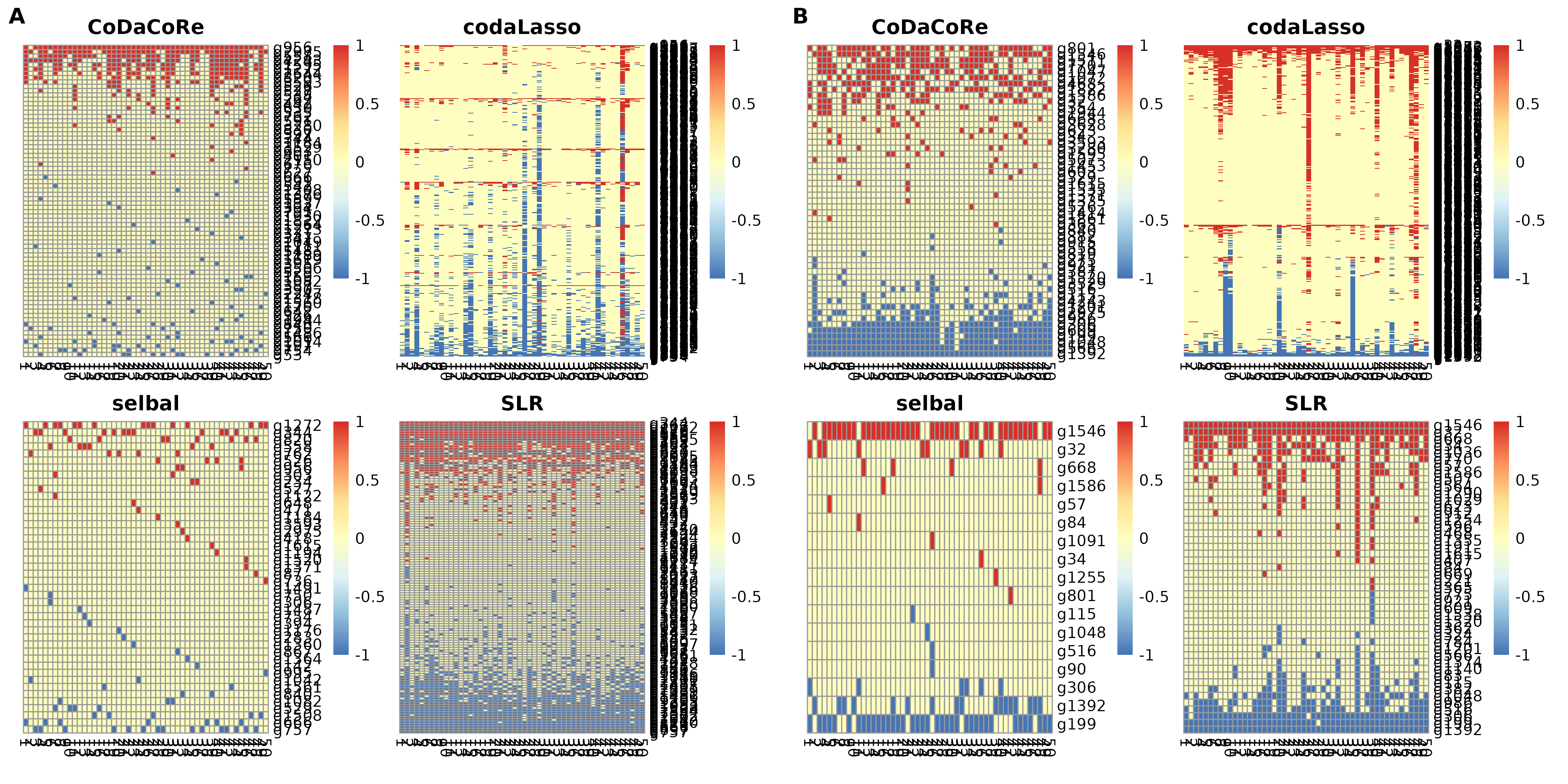}
    \caption{Heat maps of the estimated partitions from 50 replicates in the simulations based on the IBD data set when $d=400$ under Scenario I (A) and Scenario II (B). Rows indicate variables and columns indicate replicates. Only variables that were selected at least once are included.}
    \label{fig:UCsimheatmap}
\end{figure}

Figure~\ref{fig:UCsimheatmap} presents heat maps of the estimated partitions from 50 replicates when $d=400$, providing another perspective on the variable selection performance. In Scenario I (Figure~\ref{fig:UCsimheatmap}A), selbal exhibits high inconsistency in selected variables across replicates. Due to correlations among active features, once selbal identifies a log ratio that is associated with the response, it may exclude other variables that individually contribute little to AUC but collectively provide a significant improvement. As a result, it tends to prioritize different log ratios in each replicate, which would explain its low sensitivity. This pattern remains the same even when $d=50$ (Figure~\ref{fig:UCsimheatmap50}). While {\codacore} and codaLasso show greater stability than selbal, their stability is still much lower than that of SLR. In Scenario II (Figure~\ref{fig:UCsimheatmap}B), selbal consistently selects two variables. Both {\codacore} and codaLasso also demonstrate more stable variable selection in this scenario compared to Scenario I. SLR, however, selects slightly fewer consistent variables in Scenario II than in Scenario I, reflecting its lower sensitivity in this scenario. Across both scenarios, codaLasso exhibits high variability, with some replicates selecting overly dense models. Moreover, codaLasso can assign the same taxon a positive association with the response in one replicate but a negative association in another. In lower-dimensional settings, both {\codacore} and codaLasso display high variability in variable selection, with {\codacore} also exhibiting inconsistency in sign assignment (Figure~\ref{fig:UCsimheatmap50}).

All above results were based on tuning parameters selected by the 1SE rule. When we used the MIN rule to select the tuning parameters, which generally results in more variables selected by each method, the relative performance of these methods remains the same. We also observed similar patterns in the simulations based on the CRC data set (Figures~\ref{fig:CRCsim}-\ref{fig:CRCsimheatmap375}). {In summary, our simulations confirm the impact of correlation and dimensionality on accuracy in variable selection. Selbal tends to be overly conservative in variable selection, while codaLasso tends to select too many variables. The new method SLR demonstrates the best overall performance. }

\subsection{Analysis of IBD data}

We next applied these methods to the IBD data \citep{franzosa2019gut}. Recall this data set has independent discovery and validation cohort. We used the discovery cohort as the training data and the validation cohort as the test data. Because CV requires random data partition and this randomness may impact the choice of the tuning parameter, we thus performed 20 different CV runs to evaluate the consistency in variable selection. 

 \begin{figure}[h]
     \centering
    \includegraphics[width=0.99\linewidth]{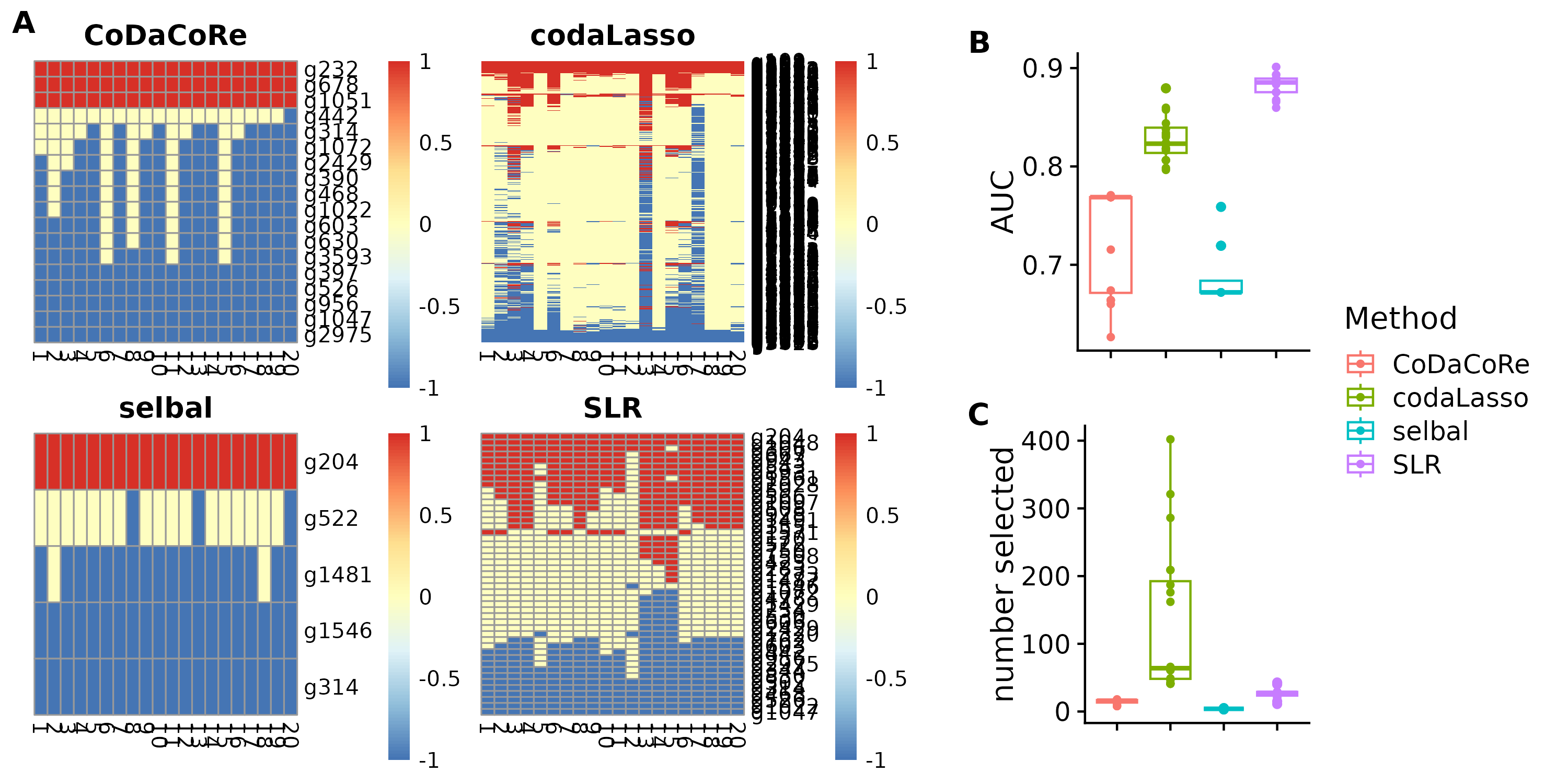}
     \caption{Results from analysis of the IBD data set \citep{franzosa2019gut} over 20 CV runs. (A) Heat maps of the estimated partitions by each method with rows indicating features and columns indicating runs. (B) AUC. (C) Number of selected features.}
     \label{fig:UC}
 \end{figure}

Figure~\ref{fig:UC} presents the heat maps of the estimated partitions along with boxplots of the classification accuracy and the number of selected features. SLR has the highest AUC and the median number of variables it selects is 27. Selbal and {\codacore} has lower AUC than SLR and codaLasso, because they both select fewer number of variables (a median of 4 for selbal and 16 for \codacore). The variables selected by {\codacore}, selbal, and SLR are largely consistent among themselves. However, codaLasso's selection shows substantial variability. This indicates that there is large variance with the optimal tuning parameter identified by CV.  

We further compared the most parsimonious model from each method. Table~\ref{tab:UC} presents the sparsest partition inferred by each method in the IBD data set, with the corresponding AUC being 0.66 for {\codacore}, 0.81 for codaLasso, 0.76 for selbal and 0.9 for SLR. 
There is no overlap in selected features among the four methods. Interestingly, there is also no overlap between variables selected by selbal and {\codacore}. The three variables selected by selbal are all identified by SLR, while SLR shares 2 variables with {\codacore}. Given the higher AUC of SLR over both selbal and {\codacore}, this could suggest that selbal and {\codacore} may have identified complementary biomarkers for UC. CodaLasso finds the most number of variables, with 29 not selected by any of the other methods. Four genera were identified by three different methods, including SLR. 
The genus {\it Mycobacterium} has a member species {\it Mycobacterium avium subspecies paratuberculosis}, which is known as the sole etiologic agent of both UC and CD \citep{pierce2010ulcerative}. {\it Fusicatenibacter} is known to produce butyrate, a short-chain fatty acid that is important for gut health. {\it Fusicatenibacter} has been found to be associated with quiescent CD \citep{chen2024enhanced}, cardiovascular disease \citep{chen2023alterations} and hepatocarcinoma \citep{huo2023altered}. The genus {\it UBA11774} is an uncultivated Lachnospiraceae bacterium and {\it Ventricola} is from the family CAG-74. Both {\it UBA11774} and {\it Ventricola} are relatively underexplored in the literature and could provide new directions for understanding UC.

\begin{table}[!ht]
\centering
\caption{The partition in the most parsimonious model by each method in the IBD data set.}\label{tab:UC} 
\resizebox{9cm}{!}{
\begin{tabular}{l|c|c|c|c}
 & CoDaCoRe & codaLasso & selbal & SLR \\ 
  \hline\hline
g\_\_UBA1685 & 1 & 1 & 0 & 0 \\ 
  g\_\_Duodenibacillus & 1 & 1 & 0 & 0 \\ 
  g\_\_Merdicola & 1 & 1 & 0 & 0 \\ 
  g\_\_Baileyella & -1 & -1 & 0 & 0 \\ 
  g\_\_Ventricola & -1 & -1 & 0 & -1 \\ 
  g\_\_Fimadaptatus & -1 & 0 & 0 & 0 \\ 
  g\_\_Fusicatenibacter & -1 & -1 & 0 & -1 \\ 
  g\_\_Avimonas\_A & -1 & -1 & 0 & 0 \\ 
  g\_\_Ruminiclostridium\_E & 0 & 1 & 0 & 0 \\ 
  g\_\_Romboutsia & 0 & 1 & 0 & 0 \\ 
  g\_\_Fournierella & 0 & 1 & 0 & 0 \\ 
  g\_\_Mycobacterium & 0 & 1 & 1 & 1 \\ 
  g\_\_Flavonifractor & 0 & 1 & 0 & 1 \\ 
  g\_\_Bifidobacterium & 0 & 1 & 0 & 0 \\ 
  g\_\_Pelethousia & 0 & 1 & 0 & 0 \\ 
  g\_\_Longicatena & 0 & 1 & 0 & 0 \\ 
  g\_\_Negativibacillus & 0 & 1 & 0 & 0 \\ 
  g\_\_CAZU01 & 0 & 1 & 0 & 0 \\ 
  g\_\_CAG-267 & 0 & 1 & 0 & 0 \\ 
  g\_\_Klebsiella & 0 & 1 & 0 & 0 \\ 
  g\_\_Acutalibacter & 0 & 1 & 0 & 0 \\ 
  g\_\_Scatomorpha & 0 & 1 & 0 & 0 \\ 
  g\_\_Massilistercora & 0 & 1 & 0 & 0 \\ 
  g\_\_51-20 & 0 & 1 & 0 & 0 \\ 
  g\_\_Monoglobus & 0 & 1 & 0 & 0 \\ 
  g\_\_GCA-900066495 & 0 & 1 & 0 & 0 \\ 
  g\_\_RUG705 & 0 & -1 & 0 & 0 \\ 
  g\_\_UBA738 & 0 & -1 & 0 & 0 \\ 
  g\_\_Megamonas & 0 & -1 & 0 & 0 \\ 
  g\_\_Peptococcus & 0 & -1 & 0 & 0 \\ 
  g\_\_UBA11774 & 0 & -1 & -1 & -1 \\ 
  g\_\_Parasutterella & 0 & -1 & 0 & 0 \\ 
  g\_\_KLE1615 & 0 & -1 & 0 & -1 \\ 
  g\_\_Slackia\_A & 0 & -1 & 0 & 0 \\ 
  g\_\_CAG-1427 & 0 & -1 & 0 & 0 \\ 
  g\_\_Fimivicinus & 0 & -1 & 0 & 0 \\ 
  g\_\_UBA1417 & 0 & -1 & 0 & -1 \\ 
  g\_\_UBA11524 & 0 & -1 & 0 & 0 \\ 
  g\_\_Gordonibacter & 0 & -1 & 0 & 0 \\ 
  g\_\_CAG-1031 & 0 & -1 & 0 & 0 \\ 
  g\_\_Scybalosoma & 0 & -1 & 0 & 0 \\ 
  g\_\_Allisonella & 0 & -1 & 0 & 0 \\ 
  g\_\_Limivivens & 0 & 0 & -1 & -1 \\ 
  g\_\_Sellimonas & 0 & 0 & 0 & 1 \\ 
  g\_\_JAGTTR01 & 0 & 0 & 0 & -1 \\ 
  g\_\_Copromorpha & 0 & 0 & 0 & -1 \\ 
   \hline
\end{tabular}
}
\end{table}

\subsection{Analysis of CRC data}

The CRC data set does not have separate discovery and validation cohort. In the benchmarking simulations, we randomly partitioned the data into a training and a test data set. However, that was done only once. To evaluate the stability in variable selection, we performed 20 different partitions, each time with 175 subjects as the training data and the remaining 75 as the test data. A single CV run was performed on each training data set.  

 \begin{figure}[h]
     \centering
    \includegraphics[width=0.99\linewidth]{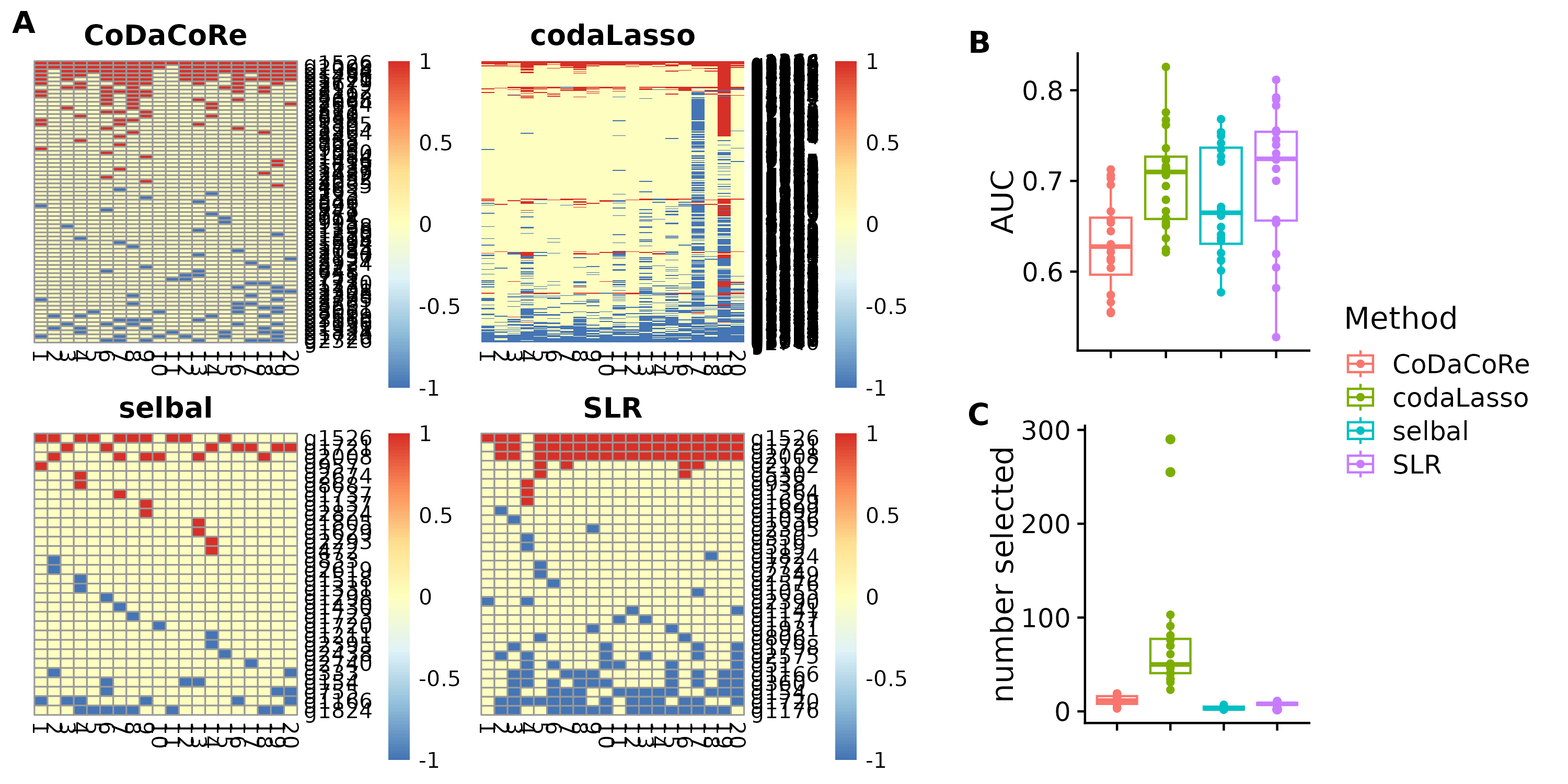}
     \caption{Results from analysis of the CRC data set \citep{yachida2019metagenomic} over 20 different training/test data splits. (A) Heat maps of the estimated partitions by each method with rows indicating features and columns indicating runs. (B) AUC. (C) Number of selected features.}
     \label{fig:CRC}
 \end{figure}

\begin{figure}
    \centering
   \includegraphics[width=0.6\linewidth]{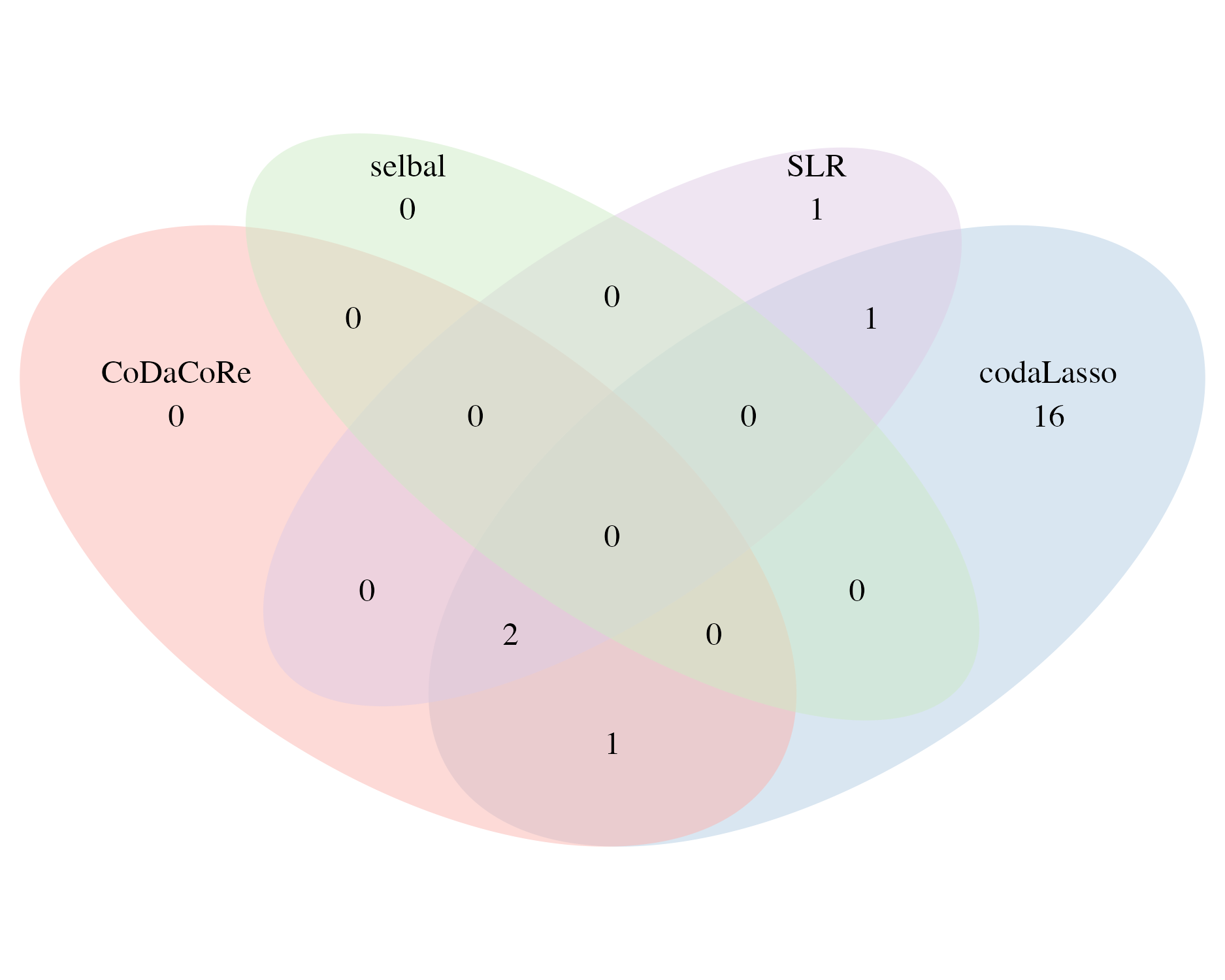}
    \caption{Venn diagram of the most frequently selected variables (at least 75\%) by the four methods in the CRC data set.}
    \label{fig:venn}
\end{figure}

Figure~\ref{fig:CRC} presents heat maps of the estimated partitions, along with boxplots of the classification accuracy and number of selected features. In terms of classification accuracy, {\codacore} falls behind the other three methods. CodaLasso has similar AUC compared to selbal and SLR, but it selects way too many variables. Due to the random data partitions, we observe more variance in selected features than in the IBD data set. However, SLR seems to show more stability in variable selection than the other methods. In fact, selbal does not select any variable more than 75\% of the time. In contrast, {\codacore} selects 3 such variables, while codaLasso selects 20. Figure~\ref{fig:venn} shows the Venn diagram of the most frequently selected variables by the four methods. SLR, {\codacore}, and codaLasso have 2 in common ({\it Fusobacterium} and {\it Parvimonas}), while codaLasso finds 16 that are not selected by any of the other methods. The genus {\it Fusobacterium} has a member species {\it F. nucleatum} that is enriched for CRC \citep{mima2016fusobacterium,kostic2013fusobacterium,flanagan2014fusobacterium}. Recent efforts focus on elucidating the exact causal mechanism underlying the association between {\it F. nucleatum} and CRC \citep{zepeda2024distinct}. {\it Fusobacterium} (marked as g2008) was selected by selbal in only 30\% of the time.  {\it P. micra}, which is the only species in the genus {\it Parvimonas}, has recently been found to be associated with immune profiles in certain CRC subtypes \citep{lowenmark2022parvimonas}. {\it Parvimonas} was selected by selbal in 50\% of the time. {\it Fusobacterium} and {\it Parvimonas} were rarely selected simultaneously by selbal (3 out of 20 runs). SLR and codaLasso both identified {\it Peptostreptococcus}, which has a species {\it P. anaerobius} that promotes colorectal carcinogenesis and modulates tumour immunity \citep{long2019peptostreptococcus}.

Finally, we provide a brief remark on the computation time. For the dimensionality in the IBD ($p=447$) and CRC data set ($p=394$), it took over 4 hours for selbal to complete both model selection and estimation using the `selbal.cv()' function of the R package {\tt selbal}, 20-30 minutes for codaLasso, about 2 minutes for SLR, and less than 1 minute for {\codacore} for each CV run. {\codacore} is a clear winner in terms of computational efficiency.

\section{Discussion}\label{sec:discussion}

 
Microbiome data are inherently compositional, meaning that all abundances are relative. Analyzing such data through multiple regression requires principled methods that are scale-invariant. Balance regression is a compelling approach as it operates in the space of log ratios, ensuring scale invariance while also providing a meaningful scalar summary of the compositional data. In this paper, we compared existing balance regression methods using data-driven simulations and two case studies. We also introduced SLR, a new method for selecting balance biomarkers in high-dimensional regression. A key aspect of the method is the preselection of taxa according to their associations with the outcome, which alleviates the effect of a large number of noisy features on the prediction model. SLR is related to supervised principal components \cite{bair2006prediction} with a key distinction of using principal balances instead of principal components to facilitate selecting interpretable balances. 
Our simulations demonstrate that SLR effectively identifies bacterial consortia, particularly when members of the consortia exhibit high correlation. This is because SLR can pool effects from multiple correlated taxa through the use of principal balance analysis and thus outperform stepwise selection methods such as selbal. 
In the two case studies, SLR identified robust and biologically meaningful bacterial signatures associated with IBD and CRC. Although our examples use microbiome data, SLR can be applied to other compositional data sets, e.g. high-throughput sequencing data from liquid biopsies \citep{gordon2022learning}.

Our findings highlight significant drawbacks of codaLasso in variable selection, including low sensitivity, low specificity, and sign inconsistency. Consistent variable selection with the Lasso often requires strong assumptions, such as the ``restricted" eigenvalue conditions \citep{van2009conditions,bickel2009simultaneous}, which are rarely met in practice. Nonetheless, codaLasso remains an effective method for predicting health outcomes. Selbal, on the other hand, is highly specific but suffers from low sensitivity. Moreover, selbal is computationally prohibitive in high-dimensional settings. This suggests that selbal may be best used in combination with exploratory techniques like SLR. While {\codacore} underperformed in our evaluations, its computational efficiency makes it an attractive choice for analyzing ultra-high-dimensional compositional data. Additionally, our study constrained {\codacore} to select only one balance for fair comparison, which may have limited its performance.

Cross-validation (CV) is widely used for model selection, but it may not be the most effective approach when the goal is robust variable selection. CV typically optimizes for prediction accuracy (e.g., MSE or AUC), which does not necessarily align with selecting stable and robust biomarkers. Our results suggest that stability selection \citep{meinshausen2010stability} could be a valuable alternative. Future research should explore additional model selection strategies to enhance variable selection performance. 

Our study differs from \citep{susin2020variable} in several key ways. First, we evaluated the impact of correlation structure on performance, while \citep{susin2020variable} focused on dimensionality and effect size. Second, our analysis considered stability in variable selection, whereas \citep{susin2020variable} relied solely on summary metrics like AUC. Third, we used CV to determine the optimal number of selected variables, while \citep{susin2020variable} maximized the proportion of explained variance. Lastly, our study focused on high-dimensional datasets with a moderate sample size, whereas \citep{susin2020variable} also analyzed a dataset with a large sample size (975) and a moderate number of taxa (48). {In low-dimensional settings where computation is not a major concern, selbal may be the preferred choice \citep{susin2020variable}.}
 
Several factors not explored in this study may also influence variable selection performance, including effect size, balance size, and the proportion of zeros in the data. We hypothesize that these factors impact all methods similarly. For example, a larger effect size likely improves the detection of active taxa across all methods, while an imbalanced numerator-to-denominator ratio in a balance may reduce selection accuracy. Further empirical studies are needed to fully understand these effects.

Log ratio-based methods require strictly positive compositional data, necessitating zero replacement before applying these methods. However, there is no consensus on how to handle zeros in microbiome studies \citep{silverman2018naught}. Some zeros reflect biological absence, while others result from limited sequencing depth. Future research should compare balance regression methods to those that do not require zero imputation \citep{shi2022high}.

\section{Data and Code Availability} 

All code needed to reproduce the results in the simulations and case studies is available at \url{https://github.com/drjingma/LogRatioReg}. SLR is available as an R package at \url{https://github.com/drjingma/slr}. 
\bibliographystyle{unsrtnat}









\section*{Acknowledgements}
This work is supported by NIH grant GM145772.

\section*{Conflict of Interest}
The authors report no conflict of interest.

\bibliography{refs}

\newpage
\section*{Supporting Information}
\setcounter{figure}{0}    
\renewcommand{\thefigure}{S\arabic{figure}}

\begin{figure}[!h]
    \centering
    \includegraphics[width=0.95\linewidth]{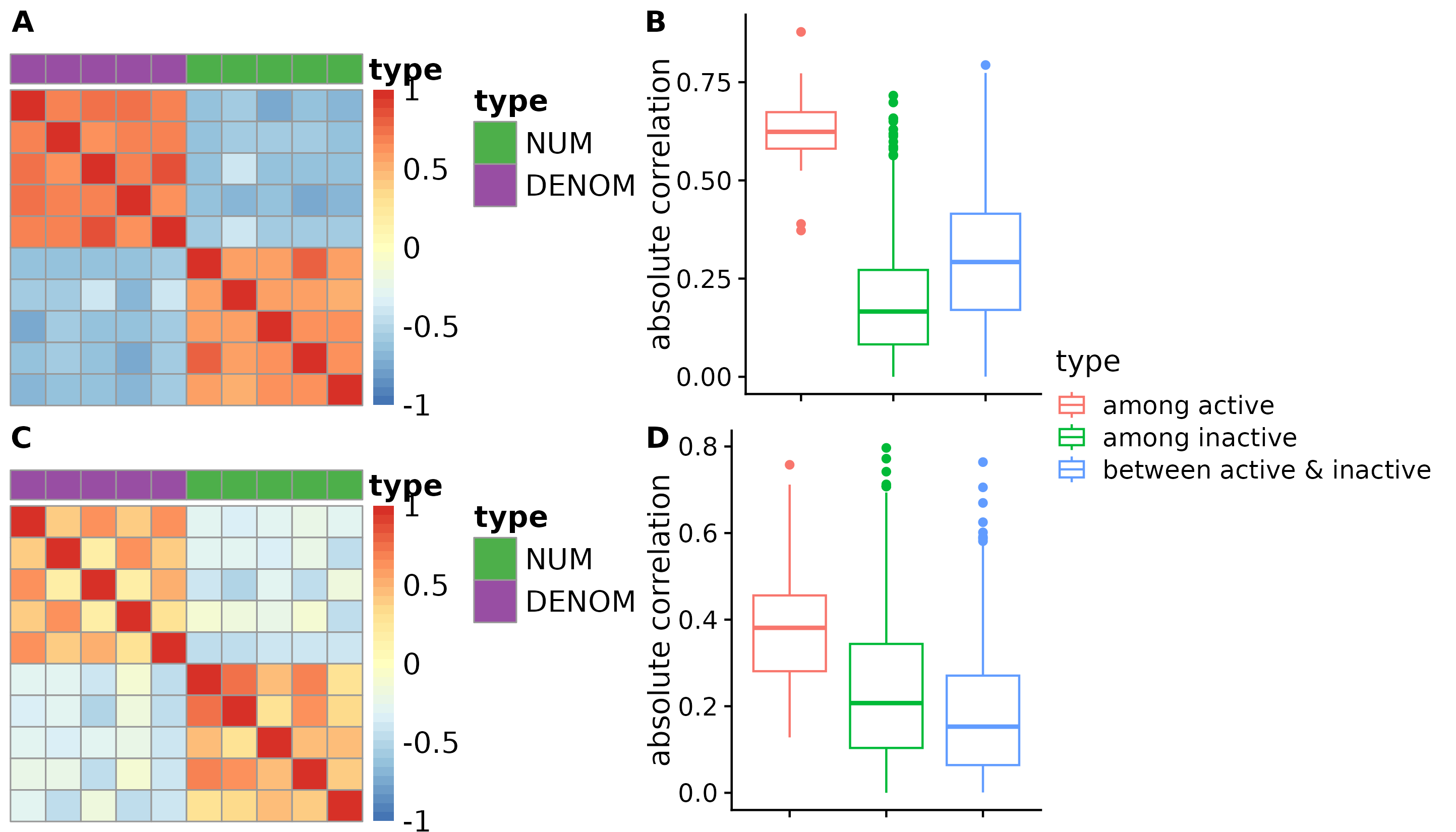}
    \caption{Correlation in the simulations based on the IBD data set ($d=50$). (A) Heat map of the correlation among active features in Scenario I. (B) Boxplots of the absolute correlation coefficients among different sets of features in Scenario I. (C) Heat map of the correlation among active features in Scenario II. (D) Boxplots of the absolute correlation coefficients among different sets of features in Scenario II.}
    \label{fig:correlation:UC50}
\end{figure}

\begin{figure}[!h]
    \centering
    \includegraphics[width=\linewidth]{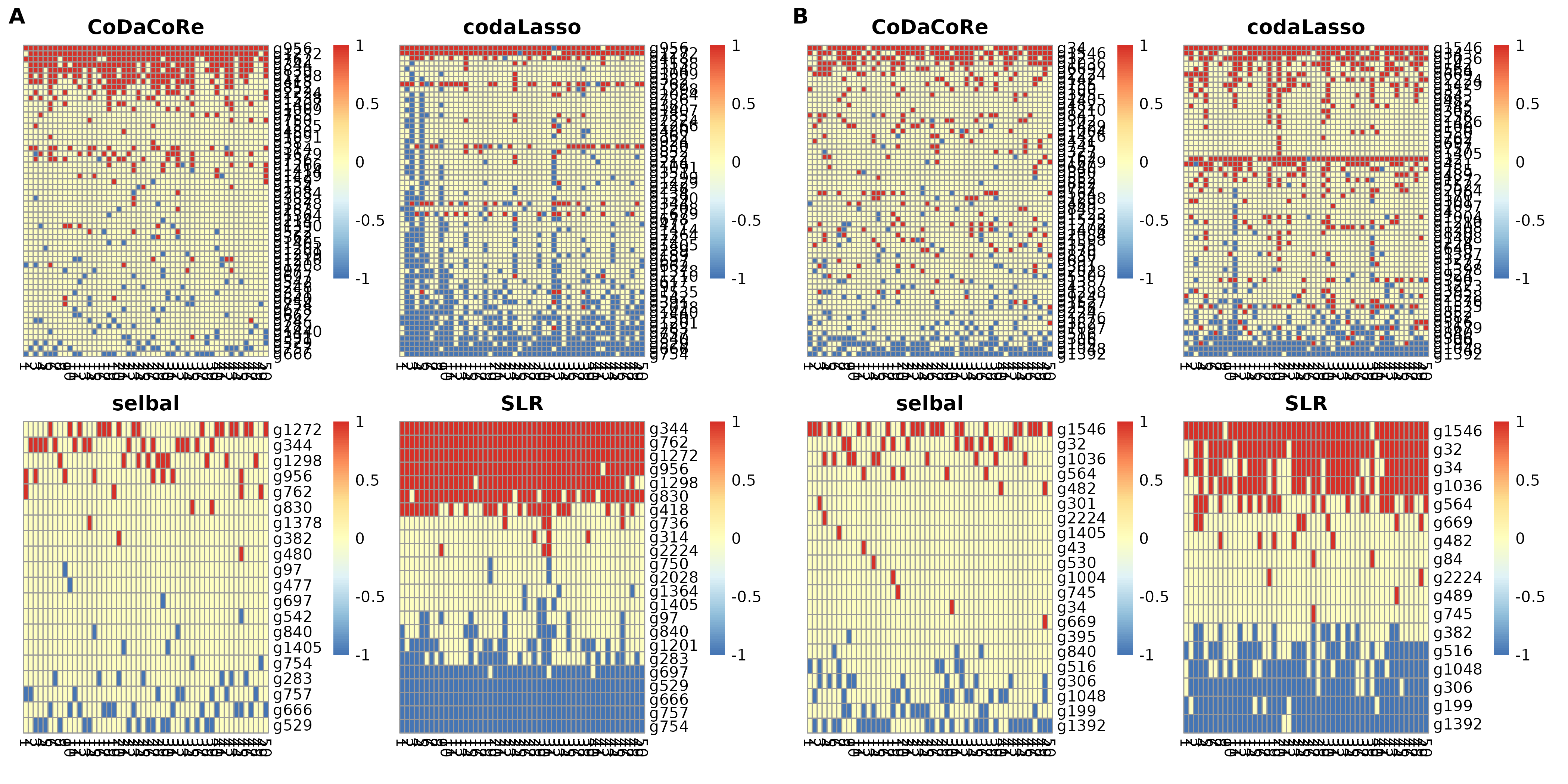}
    \caption{Heat maps of the estimated partitions from 50 replicates in the simulations based on the IBD data set example when $d=50$ under Scenario I (A) and Scenario II (B). Rows indicate variables and columns indicate replicates. Only variables that were selected at least once are included.}
    \label{fig:UCsimheatmap50}
\end{figure}

\begin{figure}[!h]
    \centering
    \includegraphics[width=0.95\linewidth]{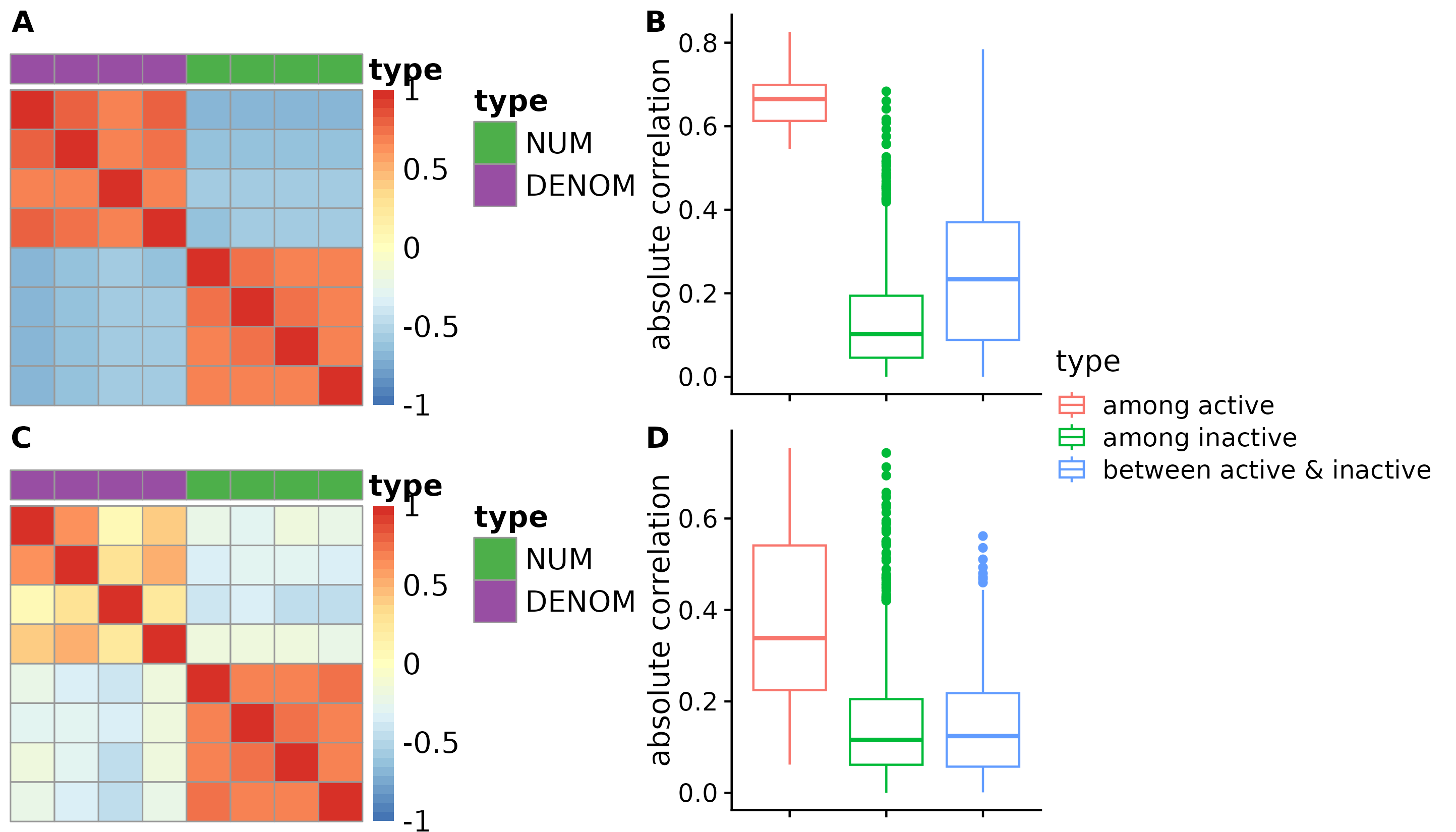}
    \caption{Correlation in the simulations based on the CRC data set ($d=50$). (A) Heat map of the correlation among active features in Scenario I. (B) Boxplots of the absolute correlation coefficients among different sets of features in Scenario I. (C) Heat map of the correlation among active features in Scenario II. (D) Boxplots of the absolute correlation coefficients among different sets of features in Scenario II.}
    \label{fig:correlation:CRC50}
\end{figure}

\begin{figure}[!h]
    \centering
    \includegraphics[width=0.95\linewidth]{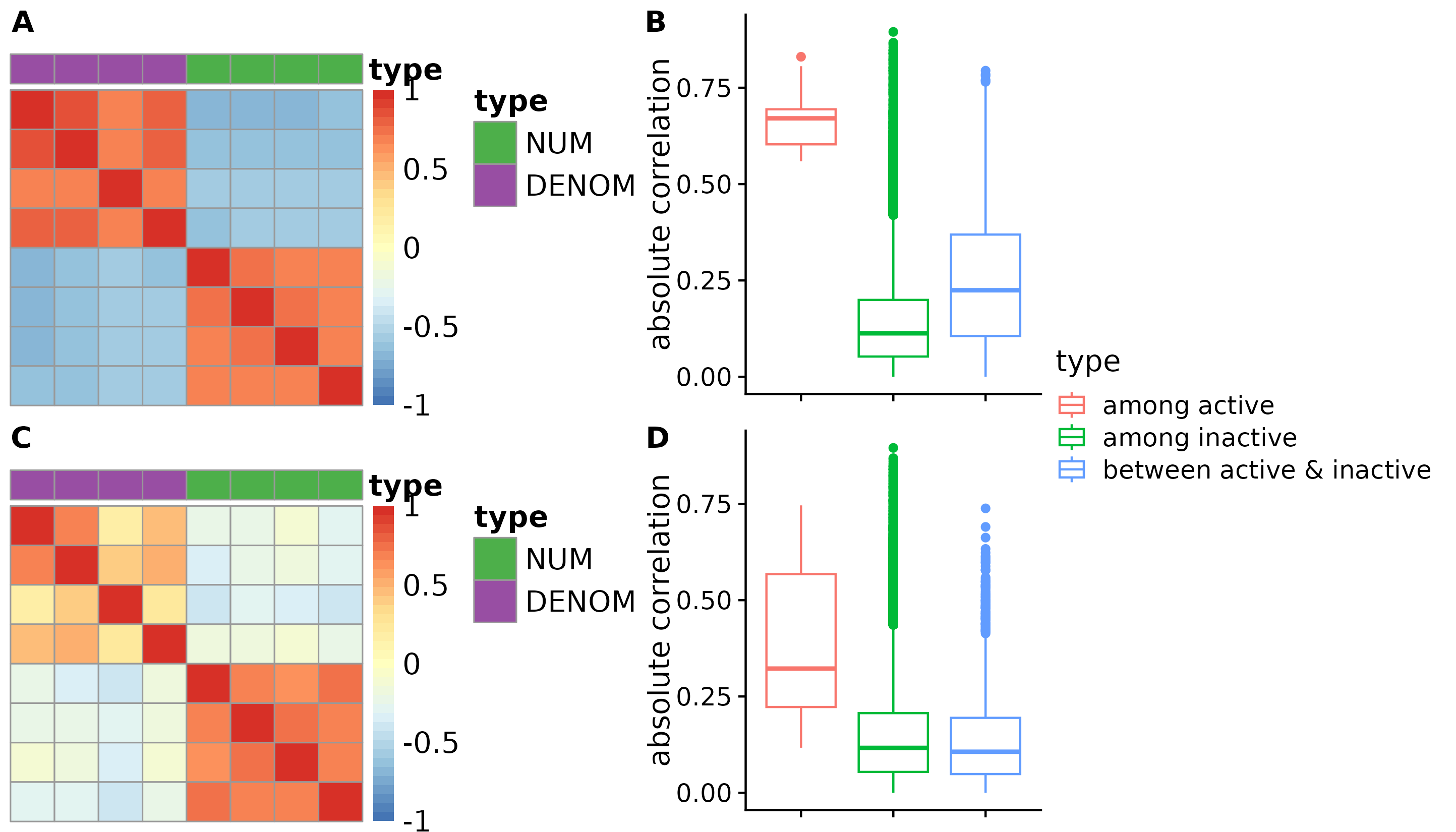}
    \caption{Correlation in the simulations based on the CRC data set ($d=375$). (A) Heat map of the correlation among active features in Scenario I. (B) Boxplots of the absolute correlation coefficients among different sets of features in Scenario I. (C) Heat map of the correlation among active features in Scenario II. (D) Boxplots of the absolute correlation coefficients among different sets of features in Scenario II.}
    \label{fig:correlation:CRC375}
\end{figure}

\begin{figure}[!h]
    \centering
    \includegraphics[width=\linewidth]{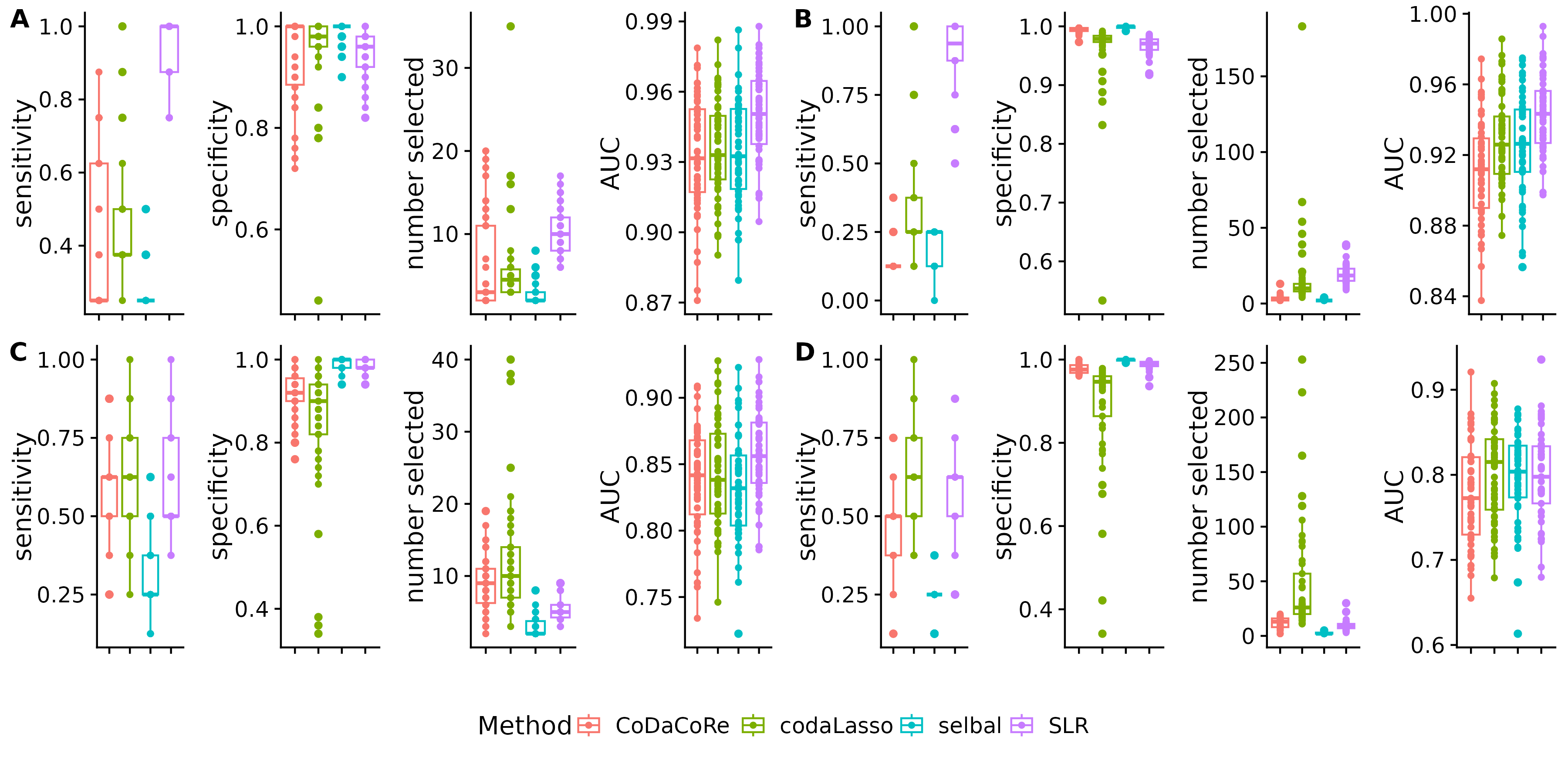} 
    \caption{Variable selection and prediction results in the simulations based on the CRC data set. Metrics used for evaluation include sensitivity, specificity, AUC, and number of variables selected. (A) Metrics in Scenario I when $d=50$. (B) Metrics in Scenario I when $d=375$. (C) Metrics in Scenario II when $d=50$. (D) Metrics in Scenario II when $d=375$.}
    \label{fig:CRCsim}
\end{figure}

\begin{figure}[!h]
    \centering
    \includegraphics[width=\linewidth]{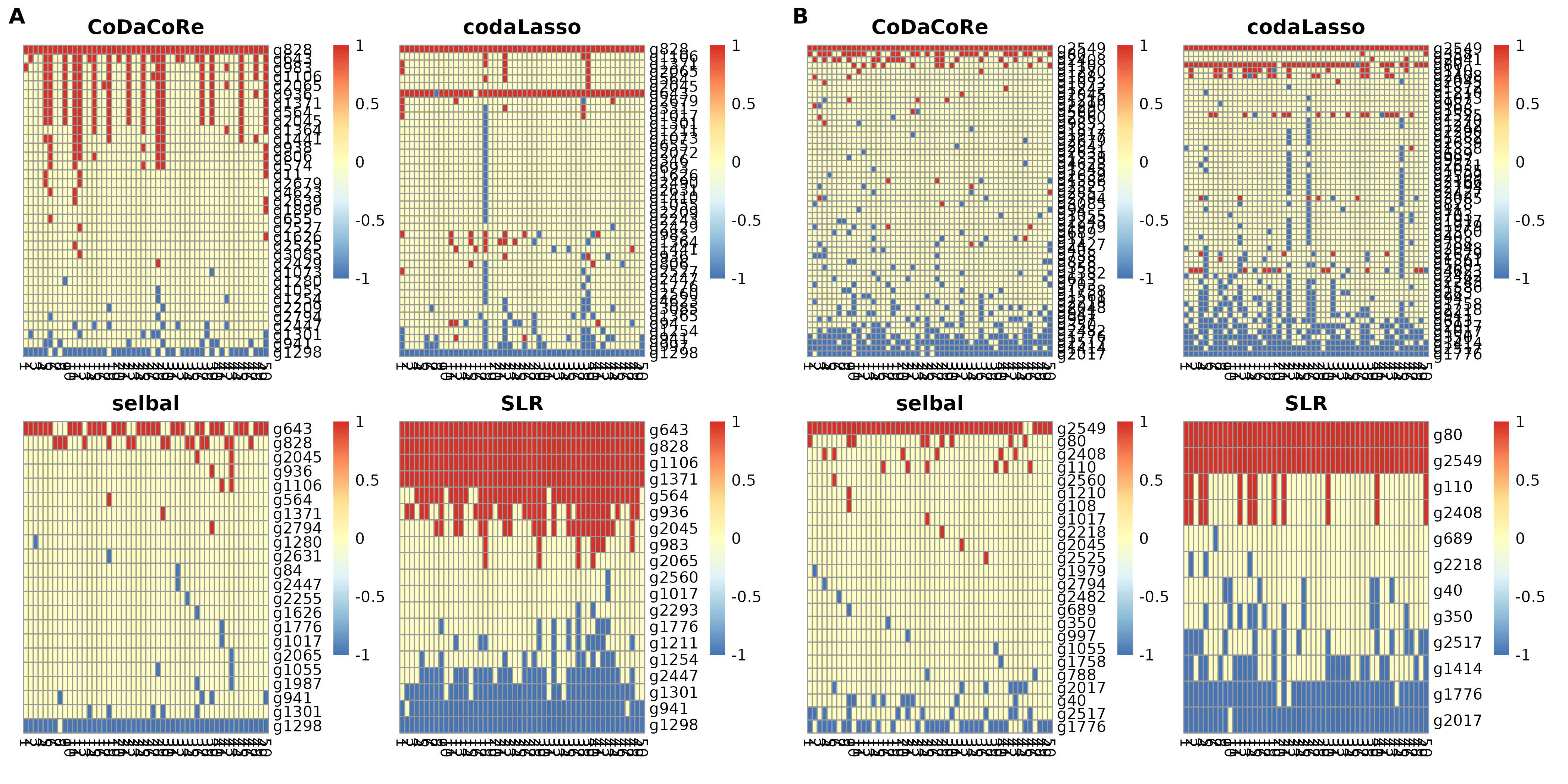}
    \caption{Heat maps of the estimated partitions from 50 replicates in the simulations based on the CRC data set when $d=50$ under Scenario I (A) and Scenario II (B). Rows indicate variables and columns indicate replicates. Only variables that were selected at least once are included.}
    \label{fig:CRCsimheatmap50}
\end{figure}

\begin{figure}[!h]
    \centering
    \includegraphics[width=\linewidth]{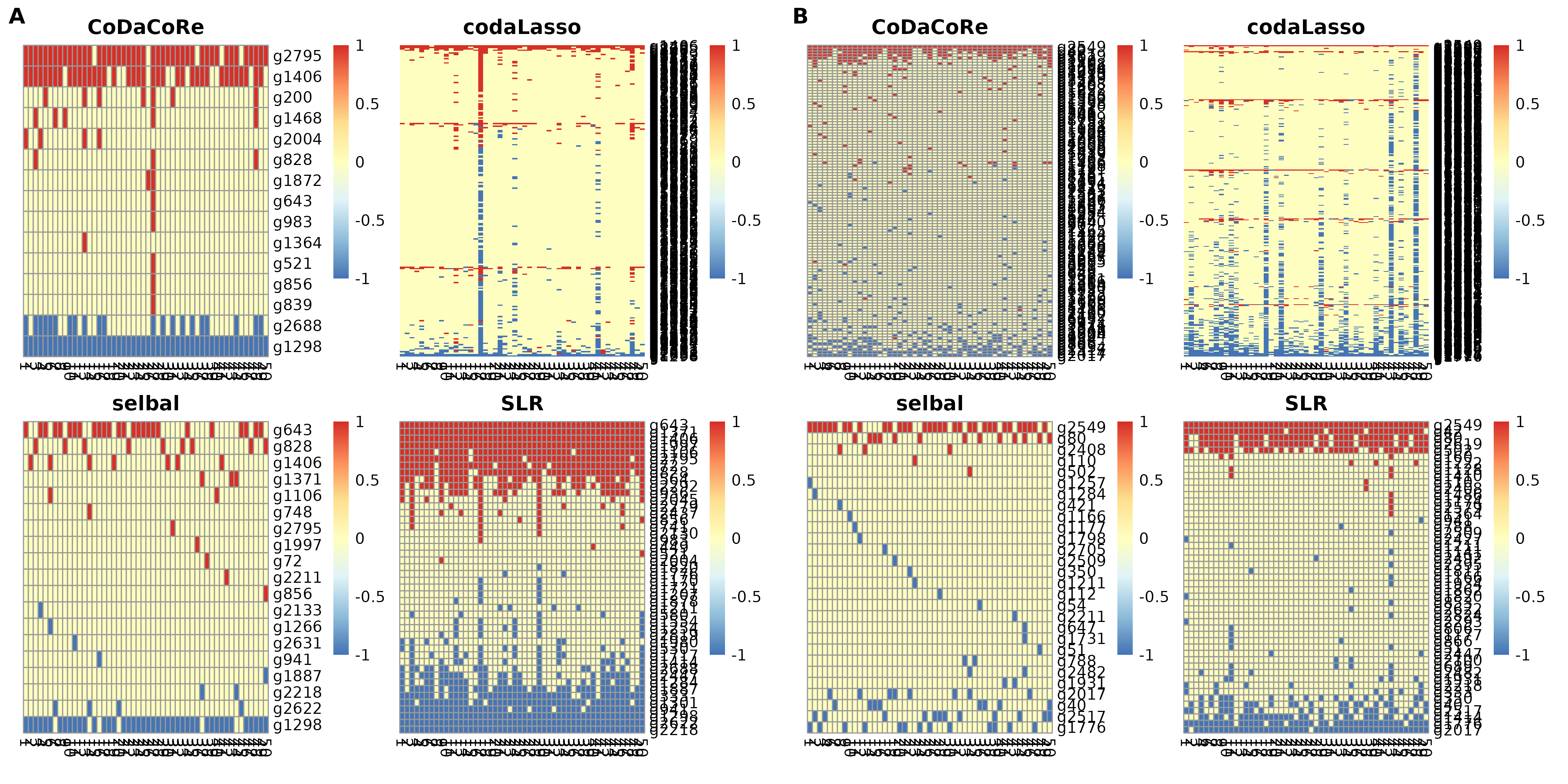}
    \caption{Heat maps of the estimated partitions from 50 replicates in the simulations based on the CRC data set when $d=375$ under Scenario I (A) and Scenario II (B). Rows indicate variables and columns indicate replicates. Only variables that were selected at least once are included.}
    \label{fig:CRCsimheatmap375}
\end{figure}

\end{document}